# Kinematic detection of the double nucleus in M31

Joris Gerssen[1], Konrad Kuijken[1]⋆ and Michael R. Merrifield[2]

[1] *Kapteyn Institute, Groningen 9700 AV, The Netherlands*
[2] *Department of Physics, University of Southampton, Highfield, UK.*

17 August 1995

**ABSTRACT**

Using a spectrum obtained under moderate ($\sim 1$ arcsecond) seeing, we show that the double nucleus in M31 produces a strong kinematic signature even though the individual components are not spatially resolved. The signature consists of a significant asymmetric wing in the stellar velocity distribution close to the center of the system. The properties of the second nucleus derived from this analysis agree closely with those measured from high-spatial resolution *Hubble Space Telescope* images. Even *Space Telescope* only has sufficient resolution to study the structure of very nearby galactic nuclei photometrically; this spectroscopic approach offers a tool for detecting structure such as multiple nuclei in a wider sample of galaxy cores.

## 1 INTRODUCTION

Because of its proximity, M31 provides us with a unique opportunity to study the detailed structure of galactic nuclei at high spatial resolution. In the 1970s, high resolution images obtained using the *Stratoscope II* telescope showed that the nucleus of M31 contains significantly asymmetric structure (Light et al. 1974). Confirmation of this unexpected result has come from *Hubble Space Telescope* (*HST*) images which show that the nucleus consists of two components separated by 0.49 arcsec (corresponding to 2 pc at the distance of M31; Lauer et al. 1993). The brighter peak, dubbed P1, appears to be unresolved and of a colour similar to the M31 bulge, whereas the second peak, P2, has a morphology which is reminiscent of the power-law cusps seen at the centres of many galactic nuclei. P2 has also been found to have a UV excess like other galactic nuclei (King et al. 1995), and it sits right at the centre of the bulge isophotes, so it seems likely that this fainter source is the "true" nucleus of M31.

Various explanations for the nuclear structure of M31 have been proposed, but so far it has proved difficult to find a scenario in which the double nucleus is not a very transient structure. Dynamical timescales at 2pc are extremely short ($\sim 10^5$ years), arguing against an out-of-equilibrium lopsidedness which is yet to phase-mix. Further, dynamical friction should have caused the orbit of any massive object to decay relatively quickly [although not as fast as naive dynamical friction arguments might suggest (Miller & Smith 1995)].

⋆ Visiting Scientist, Dept. of Theoretical Physics, University of the Basque Country, Bilbao, Spain

Tremaine (1995) has recently suggested that P1 represents the high-density part of an eccentric disk in near-Keplerian orbit about the point mass potential of P2. Crane (1995), on the other hand, has argued that the data are consistent with the possibility that the bright P1 peak is a dynamically insignificant entity on the point of merging with the true galaxy nucleus at P2.

Unfortunately, M31 is one of the few galaxies in which a structure of this kind could have been resolved, even with the *HST*. It is therefore difficult to know whether such structures are rare, and hence plausibly explained as a coincidence, or whether they are sufficiently numerous to require a more generic explanation.

The goal of the present investigation is to test whether such structures might be detected not by resolving them spatially, but by doing so spectroscopically, from their line-of-sight stellar kinematics. To this end, we have analysed high-dispersion absorption-line spectra of M31 taken in moderate ($\sim 1$ arcsec) seeing. While these data offer no hope of a spatial detection of the two components, we will show that the signal of a two-component nucleus is readily visible in the spectra. This result therefore augurs well for a search for such components in more distant galaxies, beyond the reach of even high spatial resolution imaging instruments.

## 2 THE DATA

The data described here were obtained at the end of the night of 1993 December 10 at the Multiple Mirror Telescope using the Red Channel Spectrograph with a 1.25 arcsec wide long-slit and a 1200 lines per mm grating. The dispersion per pixel with this setup is $\sim 0.7$Å ($\sim 40$ km s$^{-1}$ at 5000Å), which adequately samples the FWHM of 1.8Å. The spatial resolution was limited by the seeing, which was $\sim 1$ arcsecond FWHM. We obtained a single 30 minute integration with the slit aligned at a position angle of 148°. Calibrating arc lamp exposures were taken before and after the galaxy spectrum.

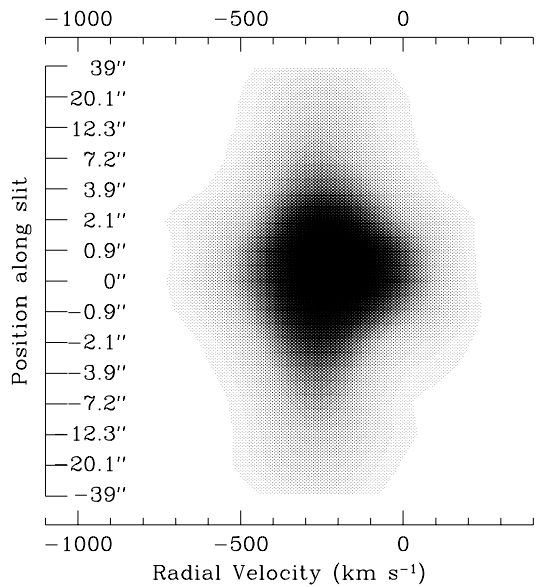

**Figure 1.** Greyscale representation of the velocity distributions in the core of M31. Darker shading indicates higher densities of stars at the given velocities and positions along the slit. Note the excess of stars at high velocities in the central arcsecond.

The long-slit spectrum was reduced to log-wavelength bins in the standard way, using IRAF packages. Adjacent spectra along the slit were then averaged to obtain a signal-to-noise ratio of at least 75 per bin, and the absorption line profiles of these co-added spectra were analysed.

## 3  ABSORPTION LINE KINEMATICS

It is now possible to derive the intrinsic shape of the broadening function of a high signal-to-noise ratio galaxy absorption line spectrum (Bender 1990, Rix & White 1992, van der Marel & Franx 1993, Gerhard 1993, Kuijken & Merrifield 1993, Saha & Williams 1994, Statler 1995). This function provides a direct measure of the line-of-sight velocity distribution (LOSVD) of the stars along that line of sight through the galaxy. To investigate the stellar kinematics in the core of M31, we have employed the unresolved gaussian decomposition (UGD) technique of Kuijken & Merrifield (1993), which returns non-negative smooth broadening functions with well-defined error bars, but does not assume a particular few-parameter description of the LOSVD.

The cumulative velocity distributions derived by the UGD method for the central regions of M31 along PA 148° are shown in Fig. 1. This slit position lies close to the minor axis of the M31 bulge, and there is little variation of the LOSVD with distance from the center outside the central few arcsec. We only show the central part of the data, since no sky-subtraction was attempted (M31 being too large for this to be possible from a single slit exposure). In the central few arcseconds of radius, a highly significant high-velocity 'blip' can be seen in these cumulative velocity curves. This feature is found independently in five adjacent spectra, and indicates an asymmetry in the central LOSVDs. It was also seen in the analysis of van der Marel et al. (1993). It is this feature in the high-velocity wing of the LOSVD which provides the signature of a secondary nucleus that we study in this paper.

## 4  IDENTIFICATION OF THE EXTRA COMPONENT

The asymmetric structure seen in Fig. 1 cannot occur in axisymmetric (or even bi-symmetric) systems, since any positive-velocity feature at a given radius should always be accompanied by a negative-velocity one, placed symmetrically with respect to the nucleus. There can therefore be no one-sided distortions to the velocity distribution in the center of a galaxy with such symmetry.

As a simple explanation for the observations (also motivated by the known photometric results) we therefore adopt a two component model. Since there is little detectable variation in the shape of the velocity distribution outside the central two arcsec along our slit position, we constructed a 'background' distribution by averaging the velocity distributions at distances between 12.3" and 20.1" on either side of the nucleus. We then made a least-squares fit simultaneously to the central five LOSVDs. As a kinematic model, we adopted the sum of the background LOSVD and an additional component which we rather arbitrarily assigned a gaussian velocity distribution whose parameters were to be determined. Both components were further required to have amplitudes which varied in a gaussian manner with position along the slit. We therefore fitted a total of eight parameters to the observed velocity distributions: three parameters for the background component's radial amplitude dependence, and five for the extra component, which is a bivariate gaussian in position and velocity.

The results of the fit are illustrated in Fig. 2 and Table 1. It is evident that the data are well described by an extra kinematic component (henceforth called K1), spatially unresolved, and kinematically offset by $+160$ km s$^{-1}$ from a background component with a spatially-invariant LOSVD and a centrally peaked density profile (henceforth K2). The brightnesses of the two components peak at slightly different positions along the slit, separated by 0.1 arcsec. The best-fit velocity dispersion of the extra component is 100 km s$^{-1}$, but simulations have shown that this number is an artefact of the smoothness imposed on the outcome of our unresolved gaussian fit to the LOSVD, and only provides an upper limit on the true dispersion of this component.

It is tempting to identify K1 and K2 directly with the photometric decomposition in to P1 and P2: in the following subsections, we assess the viability of this identification.

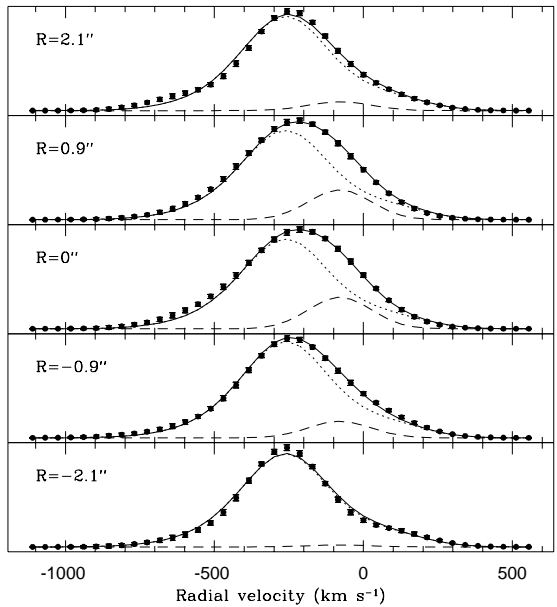

**Figure 2.** The 5 central LOSVDs with the best-fit two component model. The dotted line is K2, the dashed line is K1, and the solid line is their sum. The small error bars represent the 1-$\sigma$ range of LOSVDs, smooth on a scale of $80\,\mathrm{km\,s^{-1}}$, which are consistent with our high signal-to-noise ratio data. The indicated radii are measured with respect to the center of the CCD pixel which contained the highest flux. Vertical scales are normalized in the various panels.

**Table 1.** Best-fit parameters for the two component fit to M31's core kinematics

|  | K1 | K2 |
|---|---|---|
| Position of maximum flux | 0.4" | 0.5" |
| Total relative flux | 0.12 | 1.0 |
| Gaussian spatial dispersion | 0.95" | 2.65" |
| mean velocity | $+150\,\mathrm{km\,s^{-1}}$ | — |
| velocity dispersion | $100\,\mathrm{km\,s^{-1}}$ | — |

#### 4.0.1 Brightness

Our central five line profiles cover a rectangular region of 1.25 arcsec by 5.2 arcsec on the sky. From Bacon et al.'s (1994) fit to the central light distribution in M31, we obtain a flux ratio P1:P2 of 0.09 in this region. This value agrees very well with the relative fluxes in our kinematic K1 and K2 components: integrating the fitted contributions of both components to the line profiles in Fig. 2, we obtain a flux ratio K1:K2 of 0.12. Analysis of simulated spectra has shown that if in reality K1 has a very small velocity dispersion, its flux will be overestimated from our analysis by up to a factor two, in which case we would deduce a lower limit for the kinematic flux ratio of 0.06. These limits neatly bracket the photometric ratio.

#### 4.0.2 Position

According to the analysis of the *HST* data by Lauer et al. (1993), the photometric components are separated by 0.49" along a line with position angle 43°. Our slit was oriented at a position angle of 148°, at 75° to the P1-P2 line. Projected onto the slit, P1 and P2 are therefore separated by $0.49\cos(75°) = 0.13$ arcsec, in good agreement with the separation obtained between our fitted positions of maximum flux in K1 and K2.

#### 4.0.3 Velocity

K1 has a velocity of $+150\,\mathrm{km\,s^{-1}}$ with respect to the surrounding regions of M31. The data of Bacon et al. (1994) show a radial velocity of $+100\,\mathrm{km\,s^{-1}}$ at the peak of P1, and after correcting for seeing, these authors deduce a rotation velocity of $+160\,\mathrm{km\,s^{-1}}$ at that location, very close to our velocity offset for K1.

## 5 DISCUSSION

We have shown that it is possible to unambiguously identify the multiple components in the nucleus of M31 using kinematic observations which make no attempt to spatially resolve the components. Further, we have been able to show that the properties of the two nuclei such as their relative fluxes and velocities can be obtained reliably from such data.

From these kinematic observations, it appears that the nuclear region of M31 can be understood by postulating that P1 is simply an extra bright object in motion around the true center of the galaxy, which lies at P2. However, the dynamical friction timescale for such an arrangement is short, so (unless the object P1 has a very low mass-to-light ratio) we must be observing M31 at an unusual moment. Furthermore, any object without an unusually large mass-to-light ratio would be tidally shredded at that position, further shortening its lifetime. Perhaps a black hole, not sufficiently massive to perturb the dynamics of the center severly or produce a strong UV excess, but massive enough to accrete a substantial stellar halo of its own, may offer an explanation. Once again, though, such an object would spend most of its time at large radii, and so we are still observing M31 in an unusual state. One way around this problem has been explored by Xu & Ostriker (1994) who considered models in which a steady supply of massive black holes make their way into the nucleus.

An alternative possibility has been considered by Tremaine (1995), who has constructed a model for the central asymmetries in the light profile consisting of a lopsided disk in nearly Keplerian motion. Our data can essentially rule out such a model: by conservation of angular momentum, we know that the luminosity-weighted velocity of any filled Keplerian orbit about a stationary object must be zero. Since the putative disk would fall entirely within the slit of our observation, we would not expect to find the observed non-zero mean velocity. Non-zero velocities could result if the disk were precessing, but this scenario would require a non-Keplerian central potential which would be unlikely to support an eccentric disk over many precession periods.

A further possibility is that P1 is an M31 globular cluster which is not physically associated with the nucleus, but just lies along the same line of sight. Such a model obviates the need for a contrived dynamical model, but the very small probability of such a chance alignment renders this explanation equally unsatisfactory.

The wealth of possible explanations for the double nucleus in M31 illustrates the need for observations which will show whether such structures are rare or commonplace. Tackling this question with an imaging survey of other more distant galaxies would require challenging ultra-high resolution observations. However, we have shown that spectroscopic data can reveal the strong signature of a double nucleus even when the nuclei are unresolved, and so such observations can be used to search for more distant systems like M31 without requiring very high spatial resolution.

At very large distances (say, more than ten times the distance to M31) the spectroscopic method will also break down because of dilution of the nuclear light by scattered emission from surrounding stars. This limitation is not driven by resolution but by background noise, which makes it fundamentally different from, and less severe than, the resolution limit that applies to direct imaging of more distant galaxies.

**ACKNOWLEDGMENTS**

The data presented in this paper were obtained using the Multiple Mirror Telescope, which is a joint facility of the Smithsonian Institute and the University of Arizona. Much of the analysis was performed using IRAF, which is distributed by NOAO. MRM is supported by a PPARC Advanced Fellowship (B/94/AF/1840). We thank the referee for his helpful comments.